# A Needs Analysis Study of Amateur Astronomers For the National Virtual Observatory

Aaron Price[1]
Lou Cohen[1]
Janet Mattei[1]
Nahide Craig[2]

[1]Clinton B. Ford Astronomical Data & Research Center
American Association of Variable Star Observers
25 Birch St, Cambridge MA 02138

[2]Space Sciences Laboratory
University of California, Berkeley
7 Gauss Way
Berkeley, CA 94720-7450

**Abstract**

Through a combination of qualitative and quantitative processes, a survey was conducted of the amateur astronomy community to identify outstanding needs which the National Virtual Observatory (NVO) could fulfill. This is the final report of that project, which was conducted by The American Association of Variable Star Observers (AAVSO) on behalf of the SEGway Project at the Center for Science Educations @ Space Sciences Laboratory, UC Berkeley.

**Background**

The American Association of Variable Star Observers (AAVSO) has worked on behalf of the SEGway Project at the Center for Science Educations @ Space Sciences Laboratory, UC Berkeley, to conduct a needs analysis study of the amateur astronomy community. The goal of the study is to identify outstanding needs in the amateur community which the National Virtual Observatory (NVO) project can fulfill.

The AAVSO is a non-profit, independent organization dedicated to the study of variable stars. It was founded in 1911 and currently has a database of over 11 million variable star observations, the vast majority of which were made by amateur astronomers. The AAVSO has a rich history and extensive experience working with amateur astronomers and specifically in fostering amateur-professional collaboration.

AAVSO Director Dr. Janet Mattei headed the team assembled by the AAVSO. Dr. Mattei has been director of the AAVSO for 30 years and is in constant touch with both the amateur and professional community. Her Technical Assistant in charge of technology and public outreach, Aaron Price, designed and conducted the interviews and surveys and wrote this report. Lou Cohen, an amateur astronomer and consultant with decades of experience in corporate needs analysis projects helped with the planning process, interviewing and analyzing the results of the interviews. AAVSO Webmaster Kate Davis designed the layout of the quantitative survey. Ten members of the amateur community were interviewed and 149 responded to the quantitative survey.  Dr. Nahide Craig, Director of the Science Education Gateway program (SEGway) created the project, defined its scope and is the ultimate Project Director. This is the final report of the project.

 **The Research Process**
   In order to fully achieve the goals of this project, to identify the needs of amateur astronomers that the National Virtual Observatory can fill, we used a model of needs analysis research called *Quality Function Deployment* (QFD). QFD a well respected methodology in use for decades and is best described in the book of the same name. The author of that book is Lou Cohen, who advised us for this project. However, QFD is designed for large corporations with much greater resources so we tailored the procedure to fit the means and goals of this project.
   The amateur community was divided into five groups using a distillation process from the QFD methodology. The goal of this distillation was to come up with the smallest of number of groups possible that would encompass every amateur astronomer.  Interview subjects would then be chosen from across these groups so that we can get a cross section of the amateur community. We were able to identify five core groups using the process. The groups were: serious imagers, serious visual observers, educators, romantics, and tinkerers.

- **Romantics:** These people enjoy studying astronomy and following the latest news and research breakthroughs but do not actively observe. They prefer to read books, magazines, attend lectures, and the occasional evening star party during a major astronomical event. Many refer to themselves as "arm chair astronomers". A few of the romantics do their own astronomical research using existing online datasets.
- **Educators:** These are instructors of astronomy in the elementary, secondary, tertiary and/or adult education realms. Also included are amateur astronomers who participate in outreach activities such as giving talks, holding star parties and/or sidewalk astronomy.
- **Serious Visual Observers:** These are amateur astronomers who spend most of their time observing at the eyepiece. They enjoy pushing the limit of their equipment and observing skills. Popular projects include Messier Marathons (observing of as many Messier objects as possible in one evening), observing the Sun, Moon and planets, and attending multi-day long star parties at dark sky locations.
- **Serious Imagers:** These are advanced amateur astronomers who devote substantial amounts of their time to astronomy. They use film or CCDs to take aesthetically pleasing pictures, perform photometry, or hunt for minor planets, novae and comets. Many also get published in professional journals.
- **Tinkerers:** These are amateurs who prefer designing, building and assembling telescopes, software and other accessories as opposed to actually observing.

Resources limited us to ten interviews. We divided our subjects into these groups depending upon our estimate of their interest in the NVO web site. We chose to interview three people from the Serious Imagers category, two each from the Romantics, Educators and the Serious Visual Observers categories, and one person from the Tinkerers category. These numbers were chosen based on a predicted use of the NVO by each category of amateur astronomer.

The ten interviews were conducted in May and June of 2003. One interview was held in person at the 2003 AAVSO Spring Meeting in Tucson, Arizona. Another interview was held in person at the Desert Sun Star Party in Benson, Arizona. The remaining eight interviews were conducted via the telephone. Twenty questions were asked in each interview, which usually lasted around 40 minutes, and each interview was recorded on audiotape.

At the beginning of the interview the subject was asked questions about the NVO without being given any background information. Since one of the goals of the interview was to identify preconceived notions we did not want to bias the subject with a description of the project.

The primary goal of the interview was to identify areas of amateur astronomy where the subject feels there could be room for improvement, especially areas where the NVO can help. So the questions were focused on the subject's personal interests and experiences in amateur astronomy. Then we discussed their relationship with computers, databases, web sites and other technology.

After the interviews we listened to the tape recording three times. While listening to the tapes we wrote down every need we could identify in the subject's own words. In an ideal situation a complete transcript would be created but unfortunately resources were not available for this.

These needs were compiled into a master list. When the same need was identified from multiple sources they were consolidated into one need. This list was then divided into three sections: Core Needs, Satisfiers, and Delighters.

Once we had the master list we needed to prioritize the needs using a quantitative survey. While putting together the survey we identified two core needs that were in conflict with each other by asking the observer to choose one over the other. The first such conflict of core needs was the desire for both an easy-to-use and a powerful search engine. The second conflict involved the format of the search engine results. Some interview subjects wished for the results to be displayed on one web page while others wished for the results to be divided into many sub pages, yet for each subject this was a core need. So at the beginning of the survey we asked two either/or questions about these subjects.

The next section of the survey was a list of 24 needs. (Table 2.) The average survey subject only has the time and capability to rank between 20-25 needs. The 24 needs we chose were from the three categories (Core Needs, Satisfiers, and Delighters). Some of the needs (especially Delighters) we grouped into one general item when possible.

The goal of this relational section of the survey is to decide what is most important to the amateur community. If development assets are insufficient for fulfilling of all of these needs this information will be useful in assigning priority to the needs.

Finally, a commentary section included two open ended questions were added. They were a gateway for the subject to write any ideas, suggestions and/or needs we did not cover.

The survey was designed as an HTML page to be placed on the AAVSO web site. This allowed us to reach the entire amateur community across the country. The survey was placed online September 18 and available to the public until October 21. The AAVSO sent an e-mail announcing the survey to 360 astronomy clubs and organizations in the United States. We asked the organizations to notify their membership of the survey and to provide a link to it from their own web site. A Google search in late October found links to the survey from the web sites for The Baton Rouge Astronomical Society, Cape Cod Astronomical Society, Madison Astronomical Society, Skywatcher's Community Journal/Blog, Stellafane's WWW Site, SETI Public Mailing List, Desert Sunset Star Party WWW site, Asteroid/Comet Connection WWW site and the Yahoo Amateur Astronomy Group. We believe this is a small sample of the publicity received since Google updates their search results every 30-45 days and it had been only about a month since the announcements were sent off. Also, Google does not archive e-mail messages sent to the membership of those clubs.

In addition we e-mailed 434 members of the AAVSO Discussion Group and mentioned the survey in a publication of our CCD Views newsletter (September 24) and an announcement of new variable star charts (September 25). We also contacted astronomy news organizations asking for links and Astronomy.com (online home for Astronomy magazine) linked to the survey from their home page on October 13. In fact, they took the e-mail request and turned it into a small article about the NVO (Figure 1.).

In total we received 162 responses. We removed 13 responses because they did not complete the survey, completed it incorrectly or used the survey for other purposes (such as advertising their own survey product). The remaining 149 responses were analyzed to prioritize the needs and identify any new ones that came out of the survey. Averages were computed to establish a ranking of the needs. A standard deviation was also computed to identify the amount of consensus for each ranking. This helped identify the rankings that are more controversial than others.

**Results**

The results of the study can be summarized in two categories. First, there are the needs we were able to identify from the qualitative interviews. Second, there is the relative importance of those needs, which is established through analysis of the quantitative survey.

**Outstanding Needs of the Amateur Community**

Each interview was listened to three times: twice by Price and once by Mattei. A list of needs was pulled from each interview. Redundant entries were combined into one list encompassing 82 identified needs. (Appendix A) These needs were then grouped into 3 categories:

- **Core Needs:** These are basic core needs that the user expects from the NVO. The user will be disappointed if a need in this list is not fulfilled.

- **Satisfiers:** These are needs that can be met in a variety of levels. For example, one can barely meet a need or one can go beyond expectations. An example would be database content. Having some data meets the need but having more data than the user expects exceeds the need. Satisfiers are what users usually use to compare two products against each other. Examples: Which web search engine has better content? Which car has better gas mileage?

- **Delighters:** Delighters are bonus features that the user likely has never thought of and is quite surprised and pleased to discover in the product. (Cup holders in cars were once considered a Delighter but are now considered a Core Need.)

The *Core Needs* identified from our interview group generally involved the functionality of the product. Specifically, they were focused on the content of the database and the design of the interface. It was important to them that the content be accessible to all. Any barrier to accessing the data such as cost, web site complexity, required registration, etc. would violate this core need. Secondly, the interface needs to have all the functionality and supporting features to allow the user to understand it. Everything needs to function as advertised (i.e. "work") and manuals need to exist.

*The Satisfiers* tended to focus on the content of the database. There is a need for value-added information beyond the observational data in the database. Background information and tutorials on the data they searched for was important as was tools to further understand and possibly analyze the data. Basically we recommend the use of as much context as possible to put the data into perspective.

*The Delighters* consisted of more specific requests from the interview subjects that involved their particular area of interest. However, these bells and whistles were focused mainly on accessing the data. These are the tools that power users of the site would be interested in to save time or to help them with their *specific* observing goals. Examples include batch download of data and interfaces to other software programs. However, the list was not limited entirely to data analysis tools. Tools and materials for outreach in education and the media were frequently requested as well.

The full list of all identified needs is available in Appendix A.

**Results of the Quantitative Survey**

The results of the quantitative survey can also be divided into three categories. The first are the either/or questions involving core needs in conflict. The results illustrate the conflict quite well:

**Q: Choose the most important feature:**
    75 (50.34%): An easy-to-use search engine
    74 (49.66%): A powerful search engine

**Q: Choose the most important of either of these characteristics:**
   75 (50.34%): Search results should appear on one large page
   74 (49.66%): Search results should be organized into many sub pages

The fact that the scores were identical is interesting. However, looking at the data closely betrays this symmetry as coincidence. For example, the 75 people who preferred an easy-to-use search engine over a powerful one where not the same 75 people who preferred that the results would appear on one web page.

|  | One page | Many pages |
|---|---|---|
| Easy-to-use | 32 | 43 |
| Powerful | 43 | 31 |

**Table 1:** Breakdown of results of first two survey questions

For the relational section of the survey we computed averages and standard deviations. Lower scores reflect increased priority. The standard deviation reflects the level of consensus among the responses that the particular need fits with its location in this ranking.

| Need | Avg. score | s.d. |
|---|---|---|
| **1.** Observing charts | 8.91 | 0.920 |
| **2.** All tools should be 100% bug free | 8.94 | 1.286 |
| **3.** Include all public astronomical databases | 9.30 | 0.779 |
| **4.** Descriptions of types of data available | 10.02 | 0.811 |
| **5.** The database interface should be simple | 10.04 | 0.818 |
| **6.** Information about source of data | 10.19 | 0.210 |
| **7.** The database should interact easily with other software program | 10.87 | 0.945 |
| **8.** Links to other URLs about the observing target (OT) | 11.02 | 0.002 |
| **9.** Access to tools designed for using the data | 11.30 | 0.779 |
| **10** The web site documentation should be comprehensive | 11.56 | 1.097 |
| **11.** Planetarium point-and-click query interface | 12.05 | 1.168 |
| **12.** The web site documentation should be short and easy to read | 12.26 | 0.728 |
| **13.** Research ideas | 12.97 | 0.578 |
| **14.** There should be lots of tools for power users | 13.11 | 0.013 |
| **15.** The web site is tailored for the user's skill level | 13.21 | 1.138 |
| **16.** The site should remember me and my settings when I return | 13.26 | 1.248 |
| **17.** There should be professional guidance | 13.70 | 0.965 |
| **18.** The database interface should do it all | 13.76 | 0.028 |
| **19.** Tutorials about astronomy & math | 14.42 | 0.863 |
| **20.** The web site has a consistent look and design | 14.47 | 0.062 |
| **21.** Materials for use in public outreach activities | 14.89 | 0.568 |
| **22.** There should be a way for me to communicate with other users | 15.02 | 0.700 |
| **23.** The site should require only low bandwidth | 15.42 | 1.211 |
| **24.** The site should have a creative and exciting appearance | 18.03 | 1.631 |

**Table 2:** Results of Quantified Survey

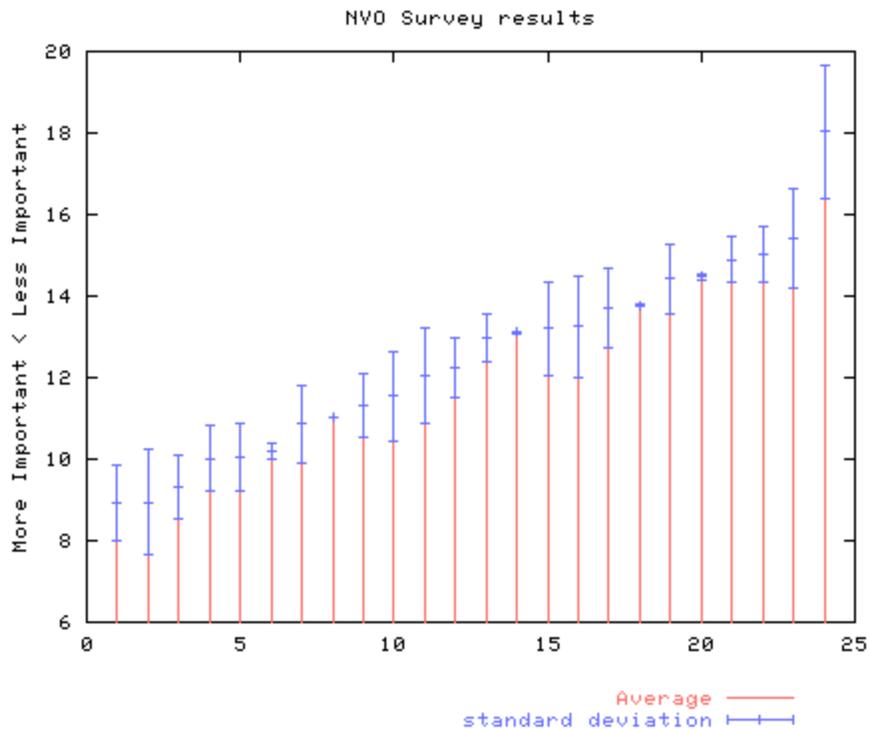

**Figure 1:** Graph of Quantified Survey Response

Finally, we added two open-ended questions to the end of the survey:

**1.** Do you have any specific features you would like to see?
**2.** Do you have any comments or suggestions?

### Recommendations

There are many outstanding needs in the amateur astronomy community that the NVO can fulfill. Fortunately most of these needs fit in with both the direction and scope of the NVO which should not be a major surprise considering the blurry line between amateur and professional astronomers. The results of this study have been augmented with the experience of the team members while working on astronomical web sites dedicated for amateur astronomers.

We recommend that a separate interface to the NVO system be designed for amateur astronomers. When designing this interface there are three core areas to focus on:

- **Functionality & Quality Control (QC):** It is vital that all aspects of the interface work. The results of our interviews were dominated by frustration with existing astronomical tools that do not work effectively and an expectation that basic functionality should work. Two ways to prevent this frustration are:

    - *Set expectations & communicate.* Accurately and clearly label the web site tools, links and capabilities. Avoid jargon and wordiness.

    - *Put the site through rigorous quality control tests.* Make sure all the links works and that the search engines and other tools function completely.

    These may seem like basic recommendations yet many organizations put out systems that are not adequately documented or tested (especially when funding and timing is an issue) and this causes frustration with amateurs, which will not be easily overcome or forgotten.

- **Web Site Content:** Put the available data in context. The amateur astronomer wants to know how the data fits into the big scheme of things, how they can observe it, what they can do with it and how the data was collected. Turn the raw data into a story.

    A tutorial regarding site functionality will be expected. In general amateurs are used to reading and following procedures, more so than the public at large. Take advantage of this uniqueness and provide an easy-to-navigate tutorial about the NVO's interface. If resources are limited, keep in mind that our survey results show that coverage of a wide variety of topics superficially is more important than focusing on a few topics in depth.

- **Database Content:** Give the amateurs access to the same data that is provided to the professionals. A significant portion of amateur astronomers are now used to collaborating with professionals and enjoying access to professional databases. They expect full access but find it difficult to manage. This fits well with the NVO core mission and could be an area where the NVO sees substantial benefit.

In a perfect world all of the needs expressed by the interview subjects should be fulfilled. If the resources do not exist to include each need in the design of the NVO, the results of the quantitative survey should be used to prioritize the needs the NVO can fulfill.

Below is a list of what we conclude are the most important specific needs which do not fall under one of the three core categories (Functionality & Quality Control, Web Site Content, Database Content) ranked in order of importance (based on results from the quantitative survey). We believe this would make for a good checklist to use during the design process.

1. Observing charts
2. The database interface should be simple
3. The database should interact easily with other software programs
4. Access to tools designed for using the data
5. Planetarium point-and-click query interface
6. Tools for power users
7. Web site customization for the user's skill level
8. The site should remember personal settings upon return
9. Professional guidance
10. Tutorials about basic astronomy & math
11. A consistent look and design
12. Materials for use in public outreach activities
13. Communication between users
14. Low bandwidth requirements
15. A creative and exciting appearance

Finally, there are some qualitative recommendations that were supplied through the quantitative survey open-ended questions. There were 59 responses of two-words or more to the open-ended questions Of the 59 responses, 38 were to the specific question "*Do you have any specific features you would like to see?*". Of those, 5 consisted of some request to have the NVO data available to other software programs. That was by far the most consistent theme in the open-ended questions. There seems to be a real desire to allow the NVO's resources to be accessed through third party software. Since there is so much third party astronomy software already available, this seems like a logical way to disseminate the NVO resources to the amateur community.

Our experience and analysis has led us to pull out a few others we felt were important enough to be emphasized:

> *"A log of recent changes or upgrades of the system."*

> *"If you take the time to design the database carefully and make sure that it is fully normalized the ability to use the data in creative and imaginative ways will be unhampered. Providing a simple interface should then be relatively easy and allowing the user to design their own custom query's should not be too far behind."*

> *"A discussion forum, possibly moderated, as an ancillary feature. No matter how well the website and user interface is designed to begin with, there will be unexpected uses, users, and difficulties; a forum is a good way to discover these, as well as encourage greater public use and participation"*

> *"ELEMENTARY Q & A or FAQ for non-scientific amateur users, i.e., poets, authors & artists who do celestial gazing for alternate reasons, but have curiosity about phenomena that may be considered very basic by astronomers"*

**Conclusion**

Amateur astronomers are a perfect fit for the NVO. They have the time, enthusiasm and ability to make use of all the unique features such a system would offer. In fact it may be that amateur astronomers become the number 1 user of such a system since they outnumber professional astronomers. It also provides the NVO with a unique community where its impact could be extreme. Because of this, particular emphasis should be placed on developing proper interfaces and support for the amateur community. Fortunately this organically falls within the guidelines, scope and goals of the NVO.

There are two basic things to always keep in mind when designing for the amateur astronomer. First, make sure the system works as advertised. Amateur astronomers are usually intelligent and successful in their own fields of work. They have high expectations. They punish failure to meet the expectations with resentment but they reward success with loyalty unheard of in other hobbies (one quarter of subscribers to Sky & Telescope magazine have held their subscription for over twenty years). Quality control should be a priority in the development of the amateur interface. Second, do not underestimate amateur astronomers. The line between professional and amateur astronomy in the United States is blurry at best. Some amateurs know the night sky better and get published in refereed journals more than some professionals. As such give them access to all the data that professionals can access. The only extra thing to consider regarding data access is that it needs to be placed in some context. Amateurs are intelligent but not (usually) professionally trained. So there is some basic jargon and knowledge that needs to be filled in.

If the NVO designs an interface that functions and is not handicapped then we believe amateur access will be a significant success and could lead to some fundamental shifts in astronomical outreach (and possibly research) strategies in the future of astronomy.

## Appendix A: List of Categorized Identified Needs

**Core Needs**
Background information about the Observing Target (OT)
A manual that is easy to access
A manual that is complete
Cheap (no cost & registration)
All links should work
Unrestricted access by all
Quality control of data
Easy navigation

All tools should function as advertised
All tools should be easier to use than existing tools
Support all public databases
Lots of links to more information about the OT
A really good search/query engine for the database
A search/query engine that is quick
Access to raw data
Up to date information
User friendly interfaces and site design
Descriptions of types of data available
Dated content
Continually updated web site

**Satisfiers**
Outreach materials that are "PR ready"
Put data in context
Tutorials on applying the data
Well-written articles about the OT
Training about how to use different file types (FITS, etc )
Ability to use without reading documentation
References to professional papers
Step-by-step how-to outlines
All answers at one site (zero clicks)
Planetarium interface
Tailored web pages for user skill level
Internal bookmarks
Common look and feel to all pages
Optimize for simplicity
Communication with others
Power tools beneath the surface
Support all coordinate systems
Image processing and star identification
Capture the excitement with multimedia & personal touches
Observing charts
Access to online tools to interpret data
Low bandwidth site
Automated batch data mining tools
All-in-one program
Faint object database
Flexible interaction with other software programs & data formats
Multiple aspects of the web site content

**Delighters**
News updates on OT
Lots of control and ability to tweak
Interdisciplinary information

Professional consultation and mentorship
Set expectations at onset
Ability to reduce data
Quick idea about what the site is about (splash screen)
Mathematical tutorials and references
Ability to reduce all the data into simple conclusions
Simple tools for common calculations
Faint multicolor stellar catalog
List of other collaborators based on their interest
Ability to add amateur data
Short answers to basic questions
Information about telescope & detector technology
Offline capabilities
Professional attitude & appearance
Materials for the media
Address misconceptions
Difficulty estimate for observing the OT
Observing tips
Observation planning tools
Voice interface
Scripting ability
Telescope control
Quality photographs
Tool to compare databases
Online telescope field of view simulator
Activities for teachers:
  - All levels from intro to advanced & short to long
  - Complete activities, no additional research needed
  - List of "a-ha's" for kids
Suggested research projects
Image processing routines
Mid level tutorials about analyzing data
Cross platform support
Materials to help teachers get support from administration
Historical information